# Gender Diversity in Ownership and Firm Innovativeness in Emerging Markets. The Mediating Roles of R&D Investments and External Capital

Vartuhi Tonoyan
California State University, Fresno
Craig School of Business
4245 N. Backer Avenue, M/S PB7
Fresno, CA 93740 US
vtonoyan@csufresno.edu

Christopher J. Boudreaux
Florida Atlantic University
College of Business
777 Glades Road, Kaye Hall 145
Boca Raton, FL 33431 US
cboudreaux@fau.edu

**Abstract.** Despite recent evidence linking gender diversity in the firm with firm innovativeness, we know little about the underlying mechanisms. Building on and extending the Upper Echelon and entrepreneurship literature, we address two lingering questions: why and how does gender diversity in firm ownership affect firm innovativeness? We use survey data collected from 7,848 owner-managers of SMEs across 29 emerging markets to test our hypotheses. Our findings demonstrate that firms with higher gender diversity in ownership are more likely to invest in R&D and rely upon a breadth of external capital, with such differentials explaining sizeable proportions of the higher likelihood of overall firm innovativeness, product and process, as well as organizational and marketing innovations exhibited by their firms. Our findings are robust to corrections for alternative measurement of focal variables, sensitivity to outliers and subsamples, and endogenous self-selection concerns.

1. Introduction

A growing body of literature has been accumulating in two management disciplines that share a common focus on gender and firm innovativeness: Upper Echelon and entrepreneurship. Despite their common interest, the two research streams have mainly developed in parallel. As a result, the findings from one area have rarely informed the other.

Studies examining the impact of gender diversity in the firm on firm innovativeness is an example of the accelerating work within the first stream, the Upper Echelon literature. Our review of the burgeoning research (summarized in the online appendix A1-A2) suggests, however, that extant research faces two limitations. First and most important, the majority of scholarship has not studied the question of why gender diversity influences firm innovativeness. Understanding "the nuts and bolts" of a managerial phenomenon, i.e., interdependencies among focal constructs and their underlying mechanisms, is a fundamental feature of theory development (Pelled, 1996). This knowledge is also essential for managers and entrepreneurs interested in designing processes to trigger specific organizational outcomes. Yet, virtually all of the investigations conducted to date (Almor et al., 2019; Biga-Diambeidou, 2019; Faems and Subramanian, 2013; Garcia-Martinez et al., 2017; Horbach and Jacob, 2018; Ruiz-Jiménez et al., 2016; Ritter-Hayashi, 2019; Talke et al., 2010; Teruel and Segarra, 2017; Xie et al., 2020), have not examined potential explanatory factors. Therefore, Pelled's (1996, p. 616) conclusion from almost 25 years ago that demographic diversity research takes "intervening processes for granted" appears to remain true today. Second, despite a growing recognition within both scholarly (Audretsch et al., 2011; Rosenbusch et al., 2011) and policy-oriented research (OECD, 2005, 2011) that innovation should not be conceived solely in terms of new products or technological process innovations, of the existing studies (reviewed in the online appendix), only Teruel and Segara (2017), Garcia-Martinez et al. (2017), and Díaz-García et al. (2013) examined additional innovative outputs such as the introduction of new marketing or organizational methods.

Although emerging scholarship from the second stream, on the "gendering" of entrepreneurship and innovation, has been more informative in unpacking some of the mechanisms explaining male-female differentials in the innovativeness of small and/or entrepreneurial firms by focusing on either the characteristics of the entrepreneur or their institutional environments (Strohmeyer and Tonoyan, 2005; Strohmeyer, Tonoyan, and Jennings, 2017; Thébaud, 2015a and 2015b), this literature has largely neglected to study the implications of gender diversity. That is, prior entrepreneurship research, with a few exceptions (Horbach and Jacob, 2018; Dai et al., 2019; Na and Shin, 2019; Ritter-Hayashi et al., 2019), has generally ignored businesses owned by the mixed-gender-teams as a focus group (Kim, 2006). This omission is unfortunate not only because it neglects a significant share of firms owned-led by both males and females (Parker, 2018), but also because it fails to answer the question of whether and why mixed-gender-owner teams differ from gender-homogeneous teams (i.e., those owned by all-males and all-females) with regard to their innovativeness.

Our research overcomes these limitations by both synthesizing insights from and extending the Upper Echelon and entrepreneurship literature. We address the above-noted gaps by theorizing why and how gender diversity in the ownership of small-and-medium-sized enterprises (SMEs) affects firm innovativeness. We delineate some of the mechanisms behind the "gender diversity in firm ownership and firm innovativeness" nexus considering the role of a firm's investments in research and development (R&D) and its financial resource endowments. The crux of our argument is that mixed-gender-owner teams will exhibit higher overall levels of firm innovativeness, and this tendency will be partially attributable to their higher likelihood of investing in R&D and relying on a breadth of external capital. Second, we focus on distinct dimensions of a firm's innovation output not yet examined in prior work — overall innovativeness, product and process innovations, and organizational and marketing innovations. Such inclusive conceptualization of firm innovativeness is essential, as it attends to various calls in both scholarly (Rosenbush et al., 2011) and policy literature (OECD, 2005, 2011) to study outputs of innovation beyond technological product and process innovation (Ayyagari et al., 2011; Strohmeyer et al., 2017; Boudreaux et al., 2019). We use the Business Environment and Enterprise Performance Survey (BEEPS) collected during 2012-2016 comprising a large sample of 7,848 randomly-selected public (i.e., stock-listed) and private (i.e., non-listed) firms from 29 emerging markets. Such a sample increases generalizability of our findings to the "understudied private firms" (Tyrowicz et al., 2019) in the literature and institutional contexts different from those of a developed economy (Jennings et al., 2013) mostly examined in prior work.

     We test our arguments using multilevel (i.e., hierarchical) linear regression to account for the nesting of firms within countries and adjusting for numerous controls at the firm and country-levels. We conduct multiple robustness checks to assess the stability and reliability of our findings. Our analyses provide reliable and robust support for our hypotheses contributing to a more comprehensive understanding of the relationship between gender diversity and firm innovativeness. Specifically, they reveal differences in the innovativeness of firms led by the mixed-gender versus gender-homogeneous owning teams, documenting that the former exhibit higher likelihoods of overall firm innovativeness, product and process innovations, as well as organizational and marketing innovations, even after correcting for the firm's selection into higher- versus lower-innovation industries. Most fundamentally, our findings unearth some of

the underlying mechanisms that have been overlooked in prior work, demonstrating the intervening roles played by a firm's investment in R&D and reliance upon a breadth of external capital. Combined, these findings offer the merit of considering alternatives to the predominant argument linking mixed-gender-owning teams with firm financial performance (for meta-analyses and reviews see, e.g., Jeong and Harrison, 2017, and Eagly, 2016). They reveal important implications for the Upper Echelon and entrepreneurship literature, as well as managers and public policy.

2. Literature review

It has been traditional for scholars in the Upper Echelon literature to study the implications of the gender diversity on corporate boards for firm financial performance, including stock market prices and volatility, return on assets, sales, and Tobin's Q (for reviews see Terjesen et al. 2009; Post and Byron, 2014; Jeong and Harrison, 2017). Grounded in management, learning theories, and social psychology, the crux of the argument is that gender diversity enhances firm financial performance due to complementarity in skills, networks, and behaviors of male and female executives. However, rigorously conducted meta-analyses (Terjesen et al. 2009; Post and Byron, 2014; Jeong and Harrison, 2017) and reviews (Eagly, 2016) concluded that beyond finding "small" and "weak" associations (Jeong and Harrison, 2017), empirical studies had not provided strong and robust support for the theory's central tenets. Some suggested female executives have small "decision latitude" on the financial performance of publicly-traded firms (Jeong and Harrison, 2017) since their appointment to corporate boards is often symbolic to "window-dress" the organization as a workplace promoting gender equality (Eagly, 2016). Also, in-group favoritism marginalizes female executives' influence on decision-making due to their "tokenism" or outgroup minority status (Kanter, 1977).

In light of these findings, scholars have suggested: (a) studying organizational contexts different from corporate boards of public firms that are more gender-inclusive providing considerable "decision latitude" to women executives and thus accentuating their impact on firm strategy (Jeong and Harrison, 2017), and (b) examining performance outcomes different from financial measures since gains of profit and productivity "are not the most appropriate place to look for diversity's benefits" (Eagly, 2013:224). Many have called to study firm innovation suggesting that gender diversity likely leads to more creativity and novel thinking within a firm, increasing its overall innovativeness (Dezsö and Ross, 2012; Eagly, 2016, 2013).

Indeed, emerging research has begun to analyze the question of whether gender diversity in the firm increases firm innovativeness, and if so, how and why. While most of the burgeoning literature (summarized in online appendix tables A1-A2) suggests it does, others do not (Biga-Diambeidou et al., 2019). It is important to note, however, that scholars employed different measures complicating the comparability of their findings. While some examined gender diversity on corporate boards (e.g., Horbach and Jacob, 2018), others explored gender diversity among employees (e.g., Østergaard et al., 2011). Many studies also adopted relatively narrow measures of innovation mostly focusing either on new product and technological process innovations (e.g., Dai et al., 2019; Ruiz-Jiménez et al., 2016), patent applications (e.g., Faems and Subramanian, 2013), or inputs to innovation process such as R&D intensity (e.g., Almor et al., 2019; Xie et al., 2020).

Unfortunately, our current understanding of why gender diversity influences firm innovation is also limited. As indicated in our review of the emerging literature (summarized in online appendix tables A1-A2), the overwhelming majority of the studies have not studied the mediating constructs linking gender diversity with firm innovativeness to explain the nature of that relationship.

## 3. Theory and hypotheses

Heeding the calls in the Upper Echelon literature to study gender-inclusive organizational contexts (Jeong and Harrison, 2017; Eagly, 2016 and 2013; Dezso and Ross, 2012), we focus on gender diversity in firm ownership. Compared to corporate boards and firm workforce, firm ownership seems a more suitable context for studying the implications of gender diversity for firm innovativeness. A firm's ownership structure aligns both incentives and actions of owners (Fama and Jensen, 1983). Hence, women's involvement in the ownership of a firm implies that they will have "decision latitude" being able to shape a firm's strategic choices (Jeong and Harrison, 2017). Also, while some studies noted potential conflicts between CEOs and in work teams that are likely to arise due to diversity of backgrounds, expertise, or values, such conflicts are less likely to occur between owner-managers because such members self-select into founding and running the firm encouraging a reciprocal exchange and collaboration (Dai et al., 2019).

We suggest that firm innovativeness should be conceptualized according to the product, process, organizational, and marketing innovation domains in which the firm has introduced something novel (Ayyagari et al., 2011; Strohmeyer et al. 2017). Such broader focus accords with Schumpeter's conceptualization of firm innovation encompassing new products, markets, production processes, and organizing methods (Schumpeter, 1934). It is also consistent with OECD guidelines (OECD, 2005, 2011) and recent calls to study novel outputs beyond new product developments or those that are primarily technological (Rosenbush et al., 2011; Strohmeyer et al., 2017; Boudreaux et al., 2019).

We posit two factors as key explanatory mechanisms underlying the relationship between gender diversity in firm ownership and firm innovativeness. The first concerns the firm's internal dynamics focusing on the decision to allocate resources to R&D. The second discusses the role of external capital.

R&D investments are a critical precursor to firm innovation (Schilling and Green, 2011). Innovations from R&D need not be technological inventions: they also include commercial discoveries, upgrades to existing products and processes, and new business models (Parker 2018: 587). Although R&D often fails to yield commercial outcomes locking firms into strategies that are difficult to change, firms across a variety of industries invest in R&D to develop new products and services and replace those threatened by substitutes (Shane 2009).

We contend that mixed-gender owning teams are more likely to invest in R&D for several reasons (Parker, 2018). First, by increasing the range of perspectives and cognitive resources available internally within the firm (Schilling and Green, 2011), gender diversity broadens the firm's knowledge base, thereby facilitating R&D undertakings (Almor et al., 2019; Miller and Triana, 2009). Second, gender diversity influences how firms recombine internal knowledge in novel ways through interaction, discussion, and mutual learning (e.g., Xie et al., 2020). Third, it expands an organization's search activities and external reach, thereby boosting the firm's absorptive capacity (Cohen and Levinthal, 1990) — a key determinant of R&D investments (Parker, 2018). Fourth, gender diversity also increases dispositional preferences for novelty and change, committing managers and employees to continuous exploration and implementation of creative ideas (Baron and Tang, 2011; Strohmeyer et al., 2017).

Although a handful of studies provide empirical support for the above-noted arguments demonstrating positive associations between various measures of gender diversity (e.g., in executive positions or R&D workforce) and R&D intensity at the firm level (Miller and Triana, 2009; Biga-Diambeidou et al., 2019; Almor et al., 2019; Xie et al., 2020), none have investigated the link between gender diversity in firm ownership, R&D investments, and firm innovation.

We posit external capital as a second intervening mechanism. External capital includes funding from banks, equity funds, micro-finance organizations, as well as suppliers, customers, family, and friends (Manolova et al., 2006). It is the "lifeblood" of firm innovation as it provides adequate capitalization vital for the development and diffusion of innovation (Gorodnichenko and Schnitzer, 2013). It also signals legitimacy, i.e., credibility and acceptance of the business by the stakeholders (Godwin, Stevens and Brenner, 2006; Beckman, Burton, and O'Reilly, 2007). Under-capitalization creates bottlenecks at various stages of innovation development, including the inability to hire and retain qualified personnel, conduct R&D, and market products and services (Shane, 2009). Innovating firms relying too heavily on internal capital often come to discover that "capital is not enough" (Bradley et al., 2012; Boudreaux and Nikolaev, 2019). Unfortunately, liquidity constraints are often more severe in emerging markets where financial systems cannot meet firms' financial needs: A survey of owner-managers of more than 15,500 firms across 29 emerging markets indicated firms' inability to access credit at reasonable terms as one of the top three business constraints (BEEPS, 2016).

We suggest that firms led by the mixed-gender-owning teams will be more likely to rely upon external funding. First, men network predominantly with other men, "important others" in business and finance, and form weak ties with former colleagues and employers (Tonoyan et al., 2020). In contrast, women often network with other women, rely on strong ties with family (for review see Jennings and Brush, 2013), and engage in support groups to overcome traditional male dominance in their industry (Vogel et al., 2014). Such network diversity increases the likelihood of obtaining both institutional and bootstrapping financing from friends and family (Manolova et al., 2006). Second, diversity due to gender differences in professional experiences and functional backgrounds is likely to send positive "signals" to financial-resource providers about the "team's completeness" (Beckman et al., 2007:123), increasing the firm's likelihood of success and hence the perception of its investment-worthiness. Mixed-gender-owner teams have the best of gender-specific behaviors and characteristics: they exhibit both a greater "task-related diversity" (Lee and Beckman, 2019), i.e., range of diverse skills and abilities, and superior relational skills combining men's agency and goal orientation with women's communality and nurturing of relationships with various stakeholders (Eagly, 2016).

A handful of studies offer empirical support for our conjectures. Vogel et al. (2014) showed that both task-oriented and relations-oriented diversity of mixed-gender entrepreneurial teams were positively related to the hypothetical resource providers' willingness to provide external capital. Beckman et al. (2007) demonstrated that task-related diversity resulting from the top management team's demographic diversity increased the start-up's ability to attract equity capital in Silicon Valley. To the best of our knowledge, the literature has not yet considered the linkages between mixed-gender-owner teams, external capital, and firm innovativeness.

In light of the preceding discussion, we develop our hypotheses as follows:

**Hypothesis 1**: Firms with higher gender diversity in ownership will exhibit higher firm innovativeness.

**Hypothesis 2:** Higher innovativeness anticipated for firms with higher gender diversity in ownership will be partly attributable to their higher likelihood of investing in R&D.

**Hypothesis 3:** Higher innovativeness anticipated for firms with higher gender diversity in ownership will be partly attributable to their higher likelihood of relying on external capital.

## 3. Methodology

*3.1. Data and sample*

Our data come from the Business Environment and Enterprise Performance Survey (BEEPS). Conducted by the World Bank and European Bank for Research and Development during 2012-2016, it is the largest firm-level survey studying managerial perceptions of the business environment in emerging markets of the post-Soviet Union, Central-Eastern Europe, Baltic, and Asia. The data are collected through face-to-face interviews with the firm owner-managers using a simple random or randomly-stratified sampling and standardized survey instruments to ensure comparability across countries (BEEPS, 2016).

BEEPS has been featured in prominent management and economics journals (Ayyagari et al., 2011; McCann and Bahl, 2017). It is well suited for testing our hypotheses for several reasons. First and most important, it is the only firm-level data containing questions pertaining to our dependent variable, focal independent variable, and proposed mediators across 29 countries. Second, it provides rich information on firm characteristics (e.g., size, age, industry sector, and establishment type) likely to influence firm innovation, which is required to test the net effects of gender diversity on firm innovativeness. A final advantage is its multi-country design. Scholars have called for cross-country studies of this nature to enhance our understanding of what determines firm innovation across emerging markets (Bradley et al., 2012; Jennings et al., 2013; Boudreaux et al., 2019). Our final sample —after removing missing observations— consists of 7,848 cross-sectional firm observations from 29 emerging markets.

*3.2. Measures*

*3.2.1. Dependent variables*

We measure firm innovativeness, our dependent variable, several different ways. Our first measure, *overall firm innovativeness*, captures firm innovation according to the dimension of width —i.e., product, process, organizational, and marketing areas— in which the firm has introduced something new (Ayygari et al., 2011; Strohmeyer et al., 2017). To measure $I_{overall}$, we created a formative index as follows:

$$I_{overall} = i_{product} + i_{process} + i_{marketing} + i_{organizational}$$

The index's four constituent items were derived from the survey questions about whether the firm had introduced any of the following initiatives during the last three years: (1) "new or significantly improved products or services" (excluding a "simple resale of new goods purchased from others and changes of a solely aesthetic nature") (product innovation); (2) "new or significantly improved methods for the production or supply of products or services" (process innovation); (3) "new or significantly improved organizational or management practices or structures" (organizational innovation), and (4) "new or significantly improved marketing methods" (marketing innovation). Each was coded dichotomously as 1=yes, 0=no. The index thus ranges from 0 to 4.

Our second measure is *product and process innovations*. Unlike many intermediary input indicators of firm innovation (such as patents or R&D intensity), new products and processes capture the commercial value of a firm's novel offerings across various industries (Audretsch et al., 2011). Its two constituent items include new product development and process innovations (each coded 1 if the firm had introduced them, 0 if not). The index thus ranges from 0 to 2.

Our third measure of a firm's innovativeness is *organizational and marketing innovations*. An organizational restructuring consists of changes in business practices, quality and human resource management, and external relations; marketing innovation includes changes to design or packaging, product promotion, placement, and pricing (OECD, 2005, 2011). Both innovation types are critical for SME competitiveness, often substituting rather than complementing product and process innovations (Audretsch et al., 2011). Its two constituent items include organizational and marketing innovations (each coded 1 if the firm had introduced these, 0 if not). The index ranges to 0 to 2.

*3.2.2. Independent and mediating variables*

Following prior literature (Talke et al., 2010; Triana et al., 2019; Xie et al., 2020; Zhang, 2020), we measure our independent variable, *gender diversity in firm ownership*, using Blau's index of heterogeneity, 2 x (1 – $\Sigma P_i^2$), where $P_i$ denotes the percentage of the population in each gender group (Blau, 1977).[1] To create this variable, we combined two questions from the BEEPS survey questionnaire: whether there are any females "amongst the owner of the firm", and if yes, "what percentage of the firm is owned by females?" The index ranges from 0 indicating complete gender homogeneity (describing all-male and all-female-owned firms) to the highest value of 1 indicating complete gender diversity (equal ownership by males and females).

Our first mediator, a *firm's R&D investments*, is coded 1 if the firm has invested in research and development activities —either in-house or contracted with other companies (outsourced) (1=yes; 0=no).

To measure our second hypothesized mediator, a firm's breadth of external capital, *Capital$_{breadth}$*, we created a formative index as follows:

$$External\ Capital_{breadth} = i_{banks} + i_{non-banks} + i_{trade\ credit} + i_{other}$$

The four constituent items were derived from the survey questions on sources of funding the firm uses to finance its working capital: (1) "banks (private and state-owned)" (banks); (2)

---

[1] For example, if women and men owned 90% and 10% of firm ownership, respectively, the Blau index would be 0.36 (2 x (1 - ($0.9^2$ + $0.1^2$)). If each of the genders possessed 50% of the firm ownership, the Blau index would be equal to one indicating gender parity (2 x (1-($0.5^2$ + $0.5^2$)).

"non-banks financial institutions" including "microfinance institutions, credit cooperatives, credit unions or finance companies" (non-banks); (3) "purchases on credit from suppliers and advances from customers" (trade credit); and (4) "other" (moneylenders, friends, relatives, etc.)" (other). Each was coded dichotomously as 1=yes, 0=no. The index ranges from 0 to 4.

*3.2.3. Control variables*

To better isolate the effects of our hypothesized predictors in HLM, we included various controls at the firm- and country-levels. Our firm-level controls capture both contemporaneous and imprinting determinants of firm innovation. Prior work has shown that larger firms, younger firms, foreign-owned firms, private firms, exporting firms, firms with specific characteristics of the Top Management, public firms, and those exposed to international markets and knowledge spillovers are more innovative (Audretsch et al., 2011).

SME is coded 1 if the firm employed less than 250 full-time employees at the time of the survey (0 otherwise). Start-up is coded 1 if the firm was less than six years of age at the time of the survey (0 otherwise). Foreign ownership is coded 1 if private foreign individuals or organizations owned more than 50 percent of the firm (0 otherwise). State ownership is coded 1 if the government held more than 50 percent of firm ownership (0 otherwise). Firm exporting is coded 1 if direct exports comprised part of the firm revenues. Top Manager describes if the firm's Top Manager was a female (coded 1 if yes, 0 if not). Industry experience is a continuous variable measuring years of industry experience of the firm's Top Manager.

International quality certification is coded 1 if the firm had an internationally-recognized quality certificate at the time of the survey (0 if not). Technology license is coded 1 if the firm had technology licenses from a foreign-owned company ("excluding office software") at the time of the survey (0 if not). Private firm/JV is coded 1 if the firm was initially established either as "private, from the time of start-up" or "joint venture with a foreign partner(s)" (0 if it was established as a "state-owned firm", "privatization of a state-owned firm", or "private subsidiary of a formerly state-owned firm"). Part of larger business is coded 1 if the firm is part of a larger business (0 if no). Public company is coded 1 if the firm is a shareholding company with (non-) publicly traded shares (0 if a sole proprietorship, partnership, or limited partnership).

We included various country-level controls to adjust for the institutional environments that are likely to influence cross-country differences in firm innovation (Boudreaux, 2017). Although we explored many such possible controls, we retained only three in our main models to have sufficient statistical power at level two of our multilevel modeling of 29 countries (Raudenbush and Bryk, 1992). [2] We included the country's per capita GDP as the country's overall economy influences both the likelihood of firm innovation and gender diversity in the firm's ownership-management (Xie et al., 2020). The country's total expenditures on research and development (R&D) (% of GDP) is a proxy for a nation's technological prowess and absorptive capacity for innovation. We also adjusted for the country's business density, which influences firm innovation either directly by encouraging the introduction of novel products or technologies as a firm differentiation strategy or indirectly through knowledge spillovers from other firms (Audretsch et al., 2011). [3] All country indicators are lagged for two years to establish causality and standardized to minimize multicollinearity.

*3.3. Analytic techniques*

We used hierarchical linear modeling (HLM) as our primary analytic technique (Raudenbush and Bryk, 2002) to control for the intra-class correlation (ICC) that was evident and attributable to the nesting of firms within countries (ICC = 0.126, $p \leq .001$ in model 1c; ICC = 0.086, $p \leq .01$ in model 2c; and ICC=0.110, $p \leq .001$ in model 3c, see Table 2). Because of the continuous nature of our innovation measures, we used multilevel linear regression models. We examined the variance inflation factor within each model to check the potential for multicollinearity. All were well below the threshold value of 10, with a mean of only 4.94.

---

[2] To obtain our country controls, we used the World Bank's Development Indicators data, EBRD's Science, Technology, and Innovation Indicators, and the US Patent and Trademark Office's patent data.

[3] Additional country indicators that we controlled for included the country's overall gender egalitarianism, women's percentage in the labor force, political instability, corruption, and voice and accountability. Models that controlled for these measures produced similar results to those presented in the main analysis (available upon request).

We supplemented the HLM results with the KHB mediation model, a rigorous approach for mediation analysis (Kohler, Karlson, and Holm, 2011) featured in prominent management journals (e.g., Tonoyan, Strohmeyer, and Jennings, 2020).[4]

## 4. Results

### *4.1. Key descriptive findings*

Table 1 reports descriptive statistics and correlations. Of the particular interest are the overall means for our innovation measures. The means for the overall innovativeness (0.87), product and process innovations (0.43), and organizational and marketing innovations (0.44) are well below these variables' theoretical maximums of 4.00, 2.00, and 2.00, respectively, suggesting that a large share of firms in our sample have not exhibited overall firm innovativeness or introduced product and process and/or organizational and marketing innovations. This finding is consistent with evidence that most SMEs are not very innovative in practice (Ruef, 2010; Thébaud, 2015a, b; Strohmeyer et al., 2017).

[Insert Table 1]

The means for our focal variables are also revealing. Approximately 11 percent of the firms invested in R&D. The mean for the breadth of external capital, at 0.70, is well below its theoretical maximum of 4.0, suggesting that only a small proportion of emerging-market firms used various sources of external capital. This finding is consistent with evidence on financing patterns of innovation in emerging markets (Manolova et al., 2006; Ayyagari et al., 2011).

---

[4] The KHB mediation model has several advantages over structural equation modeling (SEM). Unlike SEM, KHB mediation permits multiple mediators and does not require latent constructs (Kohler et al., 2011).

The bivariate comparisons between firms with and without gender-diverse ownership structures are also revealing. Consistent with our theorizing, the former are significantly more likely than that latter to exhibit overall firm innovativeness, develop product and process, as well as organizational and marketing innovations. The former are also more likely than the latter to allocate resources to R&D and rely on a breadth of external capital.

*4.2. HLM and mediation results*

Table 2 presents our HLM results. The baseline multilevel models 1a, 2a, and 3a examine the gross effects of gender diversity in firm ownership omitting control variables. We observe a positive and highly significant coefficient on gender diversity, our focal independent variable. Models 1b, 2b, and 3b test whether the net effects of our independent variable on the dependent variables remain positive and significant in the presence of the controls. The positive and highly significant coefficients for the gender diversity in firm ownership variable in these models demonstrate that they did. These findings strongly support Hypothesis 1. Models 1c, 2c, and 3c test the effects of our independent variable on firm innovation after introducing our focal mediators. The coefficients for gender diversity in firm ownership are positive and highly significant in models 1c and 3c and insignificant in model 2c. The coefficients for R&D investments and breadth of external capital are positive and highly significant. This pattern of findings suggests that higher innovativeness of firms with higher gender diversity in firm ownership is either partially or fully attributable to differences in likelihoods of R&D investments and reliance upon external capital. These results provide strong initial support for Hypotheses 2 and 3.

[Insert Table 2]

Table 3 presents the results from a more sophisticated KHB mediation analysis (Kohler et al., 2011). We observe an indirect effect of gender diversity on firm innovativeness through both of the hypothesized mediators. More precisely, R&D investments and breadth of external capital jointly account for a total of 37.06%, 45.43%, and 33.02% of the observed differentials in overall firm innovativeness, product and process innovations, and organizational and marketing innovations, respectively. These findings lend strong additional support for Hypotheses 2 and 3.

[Insert Table 3]

## 4.3. Robustness checks

### 4.3.1. Alternative measures of our focal constructs and mediators

We conducted numerous robustness checks to assess the stability of our results pertaining to alternative measures of our dependent, independent variables as well as mediators, which we summarize in Tables 4 and 5.

Table 4 summarizes the findings for each type of innovation as a separate dependent variable. Because of the dichotomous nature of the separate innovation measures, we used multilevel logistic regression models. The findings lend additional support for our Hypotheses 1-3, which predicted a higher likelihood of firm innovativeness for firms with higher gender diversity in firm ownership, with such differences being partially attributable to a firm's likelihood of R&D investments and reliance upon external capital. The total amount of mediation explained by our focal mediators corresponds to 45.83%, 35.85%, 28.21%, and 29.38% for new products, processes, organizational innovation, and marketing innovation, respectively.

[Insert Table 4]

In Table 5, we estimated multilevel models using alternative measures of our focal independent variable and the two mediators: a dichotomous measure of the gender diversity index coded 1 if the firm had a minimum of 59% of gender diversity in its ownership structure (model 1), an alternate dichotomous measure for a firm's R&D investments describing whether or a not the firm used human-resource-management practices likely to enhance firm innovativeness (model 2), a continuous measure including only funding from banks and non-bank financial institutions as the first alternate measure for external capital (model 3a), and a dichotomous measure capturing whether or not the firm received "any subsidies from the national, regional, local governments, or European Union" as the second alternate measure for external capital (model 3b). Our results are robust to these alternative specifications.

[Insert Table 5]

*4.3.2. Sensitivity to outliers and subsamples*

We assessed the sensitivity of our findings by re-estimating our multilevel models: excluding a small percentage of the solo owner-manager-led firms (model 4), checking for individual firm outlier cases (model 5), on subsamples of firms from higher-innovation industries (model 6) and lower-innovation industries (model 7). The findings provide strong additional evidence attesting to the robustness of our results.

*4.3.3. Selection correction*

We estimated the Heckman selection model (Heckman, 1979) to correct for the firm's potential self-selection into the decision to invest in R&D, on which we elaborate in online Appendix table A3 and its notes. The results strongly supported findings pertaining to our focal independent variable and external capital.

*4.4. Analytic extensions*

Instead of using a continuous variable to measure gender diversity in firm ownership, we created a dichotomous variable describing the "mixed-gender-owner teams" (i.e., firms owned equally by males and females and those with either male-dominated or female-dominated ownership structures) versus all-male-owned and all-female-owned firms. This analysis (models 1a-1c of online appendix Table A4) produced the same pattern of results demonstrating that the mixed-gender-owned teams (N=1,881) exhibited a higher likelihood of overall firm innovativeness than all-male-owned and all-female-owned teams combined (N=6,008), with the total amount of mediation explained by the mediators corresponding to 32.17%.

We then compared only all-male-owned teams versus only all-female-owned teams with the mixed-gender-owned teams. Our findings showed that all-male-owned teams (N=5,271) exhibited a significantly lower overall innovativeness than the mixed-gender-owned teams, with a lower likelihood of R&D investments and reliance on external capital mediating 36.16% of the relationship (models 2a-c, Table 6). Although all-female-owned firms (N=737) exhibited lower overall innovativeness than the mixed-gender-owned teams, and R&D investments and external capital were significant predictors of their firm innovation, these mediators did not explain the hypothesized differences with the mixed-gender-owner teams (see models 3a-c in online appendix Table A4).

We assessed the possible "treatment effects" (Rosenbaum and Rubin, 1983) to correct for the potential selection bias stemming from the fact the "treated" firms might differ from "non-treated" for reasons other than the treatment. We defined firms with a score in the upper 80-100[th] percentile of the Blau index as the treated, those with no gender diversity as the untreated. We matched the two groups on various observable characteristics pertaining to firm, industry, and country likely to impact firm innovativeness using one-to-one matching, caliper (0.01), and Mahalanobis algorithms. The treated firms had a higher likelihood of overall firm innovativeness than the non-treated. See Online Appendix Tables A5 and A6 for more information.

## 5. Discussion

The study's objective was to provide theoretical and empirical insights into the questions of how and why gender diversity in firm ownership influences firm innovativeness. With respect to the how question, our analysis of secondary data of 7,848 owner-managers of SMEs in 29 emerging markets indicates that firms with higher gender diversity in firm ownership are more likely to exhibit higher firm innovativeness. With respect to the why question, our findings provide strong and robust support for the theorized mediating roles played by the firm's R&D investments and reliance upon a breadth of external capital. We could demonstrate that sizeable proportions of the observed differentials in firm innovation—more precisely, 37.06% in overall firm innovativeness, 45.43% in product and process innovations, and 33.2% in organizational and marketing innovations—could be explained by our proposed mediators.

*5.1. Contributions to and implications for the Upper Echelon literature*

Our results extend the Upper Echelon literature in several ways. First, building on a small body of nascent work (e.g., Dai et al., 2019; Na and Shin, 2019; Ritter-Hayashi et al., 2019), we studied a highly relevant and important diversity construct —gender diversity in firm ownership—whereas most existing work (summarized in our online appendix) has examined the construct of gender diversity in other organizational contexts (such as the board of directors, senior management, task teams, and employees). We further examined firm innovativeness as a performance metric while most prior literature has studied the implications of gender diversity for firm financial performance (for a meta-analysis see Jeong and Harrison, 2017).

Our study is a first step in understanding some of the mechanisms underlying the "gender diversity and firm innovativeness" nexus that pertain to the firm's strategic allocation of resources into research and development as well as its financial resource endowments. To extend our single-mediation model, we invite scholars to test a unified "double-mediation" model (Kohler et al., 2011) that would combine the mediating effects of both the owners' individual characteristics, i.e., various types of skills, experiences, networks, and dispositional traits male and female owners bring to the table (e.g., Lee and Beckman, 2019), and their strategic practices on firm innovativeness. Future work could also examine other mediating channels (such as recruitment of specific personnel, establishing strategic alliances with business and government, and employing strategies preventing imitators from capturing the firm's profits from innovation) through which gender diversity in firm ownership is likely to influence firm innovativeness.

The socio-economic significance of our key finding about gender diversity sets the stage for a variety of related research questions. Scholars could investigate if ethnically-diverse and/or age-diverse owner teams are more innovative due to skill complementarity and strategic choices. Although emerging evidence suggests this might be the case (Hoogendoorn and van Praag, 2012; Brixy et al., 2020), we need more research.

Future studies could also study the linkage between gender diversity in ownership and firm growth through the mediating effects of firm innovativeness. Are firms headed by the mixed-gender-owner teams not only more innovative but also more likely to survive and grow? Studies of the potential downsides associated with too much diversity in firm ownership also hold promise. We invite researchers to test a curvilinear relationship to examine whether too little or too much diversity in firm ownership might hinder firm performance. Emerging evidence from a field experiment with student teams (that were required to start a business venture, elect CEOs, conduct meetings, produce, sell, make money, and liquidate within a year as part of an entrepreneurship class) demonstrated clear benefits of gender diversity. But, that relationship was curvilinear with the peak reached when women's share in the team became 0.55 (Hoogendoorn, Oosterbeek, and van Praag, 2013). Too much gender diversity is likely to become detrimental by increasing costs of coordination and communication, delaying team decisions, and slowing market moves. Yet, too little gender diversity might also hinder team performance by increasing team complacency with the information exchanged among the team members becoming too homogeneous and communication too easy.

*5.2 Contributions to and implications for the entrepreneurship literature*

Our study also extends and possesses implications for entrepreneurship research. Much current work views innovativeness as a function of the firm's organizational characteristics, alliances with venture capitalists, large corporations, and universities, as well as embeddedness in technology clusters, industries, and national innovation systems (Audretsch et al., 2011). This stream ignores, however, the question of who participates in innovation activity. As a result, some scholars have noted our limited understanding of how individual owners and entrepreneurs influence firm innovation (Baron and Tang, 2011; Strohmeyer et al., 2017). We contribute to the small but increasing body of work on this topic demonstrating the role played by individual-level characteristics —mixed-gender-owner teams— not much examined within existing entrepreneurship research.

We also extend nascent work by offering a more inclusive conceptualization of firm innovation that recognizes not only new products and technological process innovation but also overall firm innovativeness along with the separate individual domains (i.e., product, process, organizational, and marketing) in which the firm has introduced something novel (Ayyagari et al., 2011; Strohmeyer et al., 2017). That said, future researchers could examine the consequences of gender diversity in firm ownership for "innovation depth" or the radicalness of the firm's new offerings regardless of the individual domains where they were introduced. Radical innovations typically emerge after recombining existing ideas from different domains that previously seemed unrelated (Hargadon, 2003; Schilling and Green, 2011). As mixed-gender-owned teams are more likely to possess knowledge and experience from multiple unrelated domains, they might also be more likely to introduce radical innovation.

We also contribute to the literature on women's entrepreneurship. Although recent research started to study entrepreneurship as a "gendered process" (Brush, 1992; Jennings and Brush, 2013), only a few studies (Horbach and Jacob, 2018; Dai et al., 2019; Na and Shin, 2019; Ritter-Hayashi et al., 2019) examine mixed-gender ownership as a focus group. Yet, ignoring businesses owned by mixed-gender teams is problematic, since such omission overlooks how complements in skills, networks, and characteristics of male and female entrepreneurs shape firm-level outcomes and evaluations of resource providers (Kim, 2006; Parker, 2018). We demonstrated that a male-female partnership was more beneficial for firm innovativeness than all-male partnership because such diversity enhanced the likelihood of R&D investments and reliance upon external capital. We offered a theory suggesting that the addition of a partner to the firm ownership outside the "in" network who can bring with them skills or attributes that the other partners lack (Godwin et al., 2006) will enhance the firm innovativeness. Unexpectedly, however, our proposed mediators did not explain the gaps in the innovativeness of firms owned by all females and the mixed genders. Such a finding merits more detailed investigation in future research.

We also encourage future research on the moderating effects of a country's institutional environment: given that the degree of sex-based segregation in business ownership and innovation varies across countries (Thébaud, 2015a; Tonoyan et al., 2020), partnering with a male (or female) business owner might confer more benefits in securing critical resources and organizational legitimacy (Godwin et al., 2006) in some institutional contexts than others.

As some suggested that female owner-managers might prefer creating organizations that are more egalitarian and less hierarchical (Jennings and Brush, 2013), future research is also warranted for studying the importance of the organizational culture. Gender diversity in ownership might create climates for inclusion, fostering employees' participation in problem-solving, diversity in opinions, and encouraging exchange between the firm and its constituents (e.g., customers and regulators) (Cropley and Cropley, 2017). Such channels are likely to enhance employee creativity, firm absorptive capacity, and hence innovativeness.

## 6. Limitations and suggested directions

The results of our research are subject to a few limitations, mainly stemming from the secondary nature of our focal dataset, that should be addressed in future research. Although our results were robust to various sensitivity checks and self-selection concerns, due to the cross-sectional nature of our data, we encourage the replication of our findings with longitudinal firm data. Since our sample mostly comprised emerging markets from the post-Socialist countries, we also encourage replication of our results on samples of different countries. Finally, we focused solely upon the firm's lead owner-managers assuming that such individuals as heads of firms are those who "call the shots", i.e., make strategic decisions likely to impact firm innovativeness. But, innovation is an inclusive process often involving the input of highly-capable employees from various functional areas (such as engineering, marketing, and manufacturing) (Østergaard et al., 2011). Scholars should thus study the interplay between skill diversity of owner-managers and specialization amongst the firm employees to extend our understanding of the antecedents of firm innovation.

## 7. Conclusion

Returning to our guiding questions of how and why gender diversity in firm ownership influences firm innovativeness, we conclude with the following responses. With respect to the how questions, firms with higher gender diversity in ownership in our sample differed significantly in their degree of innovativeness, exhibiting higher likelihoods of overall innovativeness, product and process, as well as organizational and marketing innovations. As for the why question, sizeable proportions of these differentials were attributable to a higher likelihood of investing in R&D and relying upon external capital.

We would like to provide a final note about our study's implications for managers and policymakers. A managerial implication is that both male and female business owners should consider partnering with the opposite sex, as gender diversity in ownership significantly increases the odds that the firm will become more innovative along the dimensions of new products, processes, organizational practices, and marketing methods. A policy implication is that policymakers might want to focus on desegregating business ownership. Providing institutional, social, and resource-based support for increasing the share of the mixed-gender-owner teams will go a long way in enhancing firm innovativeness.

Strohmeyer, R., Tonoyan, V., Jennings, J.E., 2017. Jacks-(and Jills)-of-all-trades: On whether, how and why gender influences firm innovativeness. Journal of Business Venturing 32, 498–518. https://doi.org/10.1016/j.jbusvent.2017.07.001

Talke, K., Salomo, S., Rost, K., 2010. How top management team diversity affects innovativeness and performance via the strategic choice to focus on innovation fields. Research Policy 39, 907–918. https://doi.org/10.1016/j.respol.2010.04.001

Terjesen, S., Sealy, R., Singh, V., 2009. Women directors on corporate boards: A review and research agenda. Corporate Governance: An International Review 17, 320–337. https://doi.org/10.1111/j.1467-8683.2009.00742.x

Teruel, M., Segarra, A., 2017. The link between gender diversity and innovation: What is the role of firm size? International Review of Entrepreneurship 15, 319–340.

Thébaud, S., 2015a. Business as plan B: Institutional foundations of gender inequality in entrepreneurship across 24 industrialized countries. Administrative Science Quarterly 60, 671–711. https://doi.org/10.1177/0001839215591627

Thébaud, S., 2015b. Status beliefs and the spirit of capitalism: Accounting for gender biases in entrepreneurship and innovation. Social Forces, 94, 61-86.

Tonoyan, V., Strohmeyer, R., Jennings, J.E., 2020. Gender gaps in perceived start-up ease: Implications of sex-based labor market segregation for entrepreneurship across 22 European countries. Administrative Science Quarterly 65, 181–225.

Triana, M. del C., Richard, O.C., Su, W., 2019. Gender diversity in senior management, strategic change, and firm performance: Examining the mediating nature of strategic change in high tech firms. Research Policy 48, 1681–1693. https://doi.org/10.1016/j.respol.2019.03.013

Tyrowicz, J., Terjesen, S., Mazurek, J., 2020. All on board? New evidence on board gender diversity from a large panel of European firms. European Management Journal. https://doi.org/10.1016/j.emj.2020.01.001

Vogel, R., Puhan, T.X., Shehu, E., Kliger, D., Beese, H., 2014. Funding decisions and entrepreneurial team diversity: A field study. Journal of Economic Behavior & Organization 107, 595–613. https://doi.org/10.1016/j.jebo.2014.02.021

Xie, L., Zhou, J., Zong, Q., Lu, Q., 2020. Gender diversity in R&D teams and innovation efficiency: Role of the innovation context. Research Policy 49, 103885. https://doi.org/10.1016/j.respol.2019.103885

Zhang, L., 2020. An institutional approach to gender diversity and firm performance. Organization Science. https://doi.org/10.1287/orsc.2019.1297
29

**Table 1: Descriptive statistics and correlations**

| | | Mean | SD | Gender diverse ownership Firms with | Gender diverse ownership Firms without | Diff. stat. | 1 | 2 | 3 | 4 | 5 | 6 | 7 | 8 | 9 | 10 | 11 | 12 | 13 | 14 | 15 | 16 | 17 | 18 | 19 | 20 | 21 | 22 | 23 | 24 | 25 | 26 |
|---|---|---|---|---|---|---|---|---|---|---|---|---|---|---|---|---|---|---|---|---|---|---|---|---|---|---|---|---|---|---|---|---|
| 1 | Over'll firm innov. | 0.87 | 0.01 | 1.04 | 0.81 | t=6.66*** | 1 | | | | | | | | | | | | | | | | | | | | | | | | | |
| 2 | Prod. & proc. innov. | 0.43 | 0.01 | 0.51 | 0.41 | t=5.15*** | 0.88 | 1 | | | | | | | | | | | | | | | | | | | | | | | | |
| 3 | Org. & markt. innov. | 0.44 | 0.01 | 0.53 | 0.41 | t=6.48*** | 0.89 | 0.56 | 1 | | | | | | | | | | | | | | | | | | | | | | | |
| 4 | Product innov. | 0.24 | 0.00 | 0.28 | 0.23 | t=4.18*** | 0.75 | 0.88 | 0.45 | 1 | | | | | | | | | | | | | | | | | | | | | | |
| 5 | Process innov. | 0.19 | 0.00 | 0.23 | 0.18 | t=4.80*** | 0.77 | 0.85 | 0.51 | 0.49 | 1 | | | | | | | | | | | | | | | | | | | | | |
| 6 | Organizational innov. | 0.21 | 0.00 | 0.25 | 0.19 | t=5.70*** | 0.79 | 0.51 | 0.88 | 0.40 | 0.49 | 1 | | | | | | | | | | | | | | | | | | | | |
| 7 | Marketing innovation | 0.23 | 0.00 | 0.28 | 0.21 | t=5.76*** | 0.78 | 0.48 | 0.89 | 0.40 | 0.42 | 0.57 | 1 | | | | | | | | | | | | | | | | | | | |
| 8 | Gender-diverse owner. | 0.18 | 0.00 | n/a | n/a | n/a | 0.07 | 0.05 | 0.06 | 0.05 | 0.05 | 0.05 | 0.05 | 1 | | | | | | | | | | | | | | | | | | |
| 9 | R&D activity | 0.11 | 0.00 | 0.14 | 0.10 | t=4.91*** | 0.41 | 0.37 | 0.35 | 0.32 | 0.32 | 0.32 | 0.30 | 0.05 | 1 | | | | | | | | | | | | | | | | | |
| 10 | Breadth of extern. | 0.70 | 0.01 | 0.78 | 0.67 | t=5.10*** | 0.17 | 0.15 | 0.15 | 0.13 | 0.12 | 0.13 | 0.14 | 0.06 | 0.10 | 1 | | | | | | | | | | | | | | | | |
| 11 | SME | 0.96 | 0.00 | 0.95 | 0.96 | t=-2.09** | -0.08 | -0.05 | -0.09 | -0.04 | -0.04 | -0.10 | -0.06 | 0.00 | -0.07 | -0.05 | 1 | | | | | | | | | | | | | | | |
| 12 | Startup | 0.16 | 0.00 | 0.12 | 0.17 | t=-5.19*** | -0.05 | -0.06 | -0.04 | -0.05 | -0.05 | -0.04 | -0.03 | -0.05 | -0.04 | -0.07 | 0.05 | 1 | | | | | | | | | | | | | | |
| 13 | Top Manager: Female | 0.19 | 0.00 | 0.30 | 0.15 | t=14.16*** | 0.02 | 0.02 | 0.02 | 0.01 | 0.01 | 0.01 | 0.03 | 0.17 | -0.03 | -0.02 | 0.07 | 0.01 | 1 | | | | | | | | | | | | | |
| 14 | Years of industry exper | 17.99 | 0.12 | 19.44 | 17.54 | t=6.82*** | 0.05 | 0.05 | 0.03 | 0.05 | 0.03 | 0.04 | 0.02 | 0.08 | 0.08 | 0.03 | -0.06 | -0.27 | -0.07 | 1 | | | | | | | | | | | | |
| 15 | Foreign ownership | 0.05 | 0.00 | 0.04 | 0.05 | t=-2.46** | 0.07 | 0.05 | 0.07 | 0.05 | 0.04 | 0.06 | 0.06 | -0.04 | 0.04 | -0.02 | -0.12 | 0.02 | -0.02 | -0.04 | 1 | | | | | | | | | | | |
| 16 | State ownership | 0.01 | 0.00 | 0.01 | 0.01 | t=0.93 | -0.02 | -0.02 | -0.02 | -0.02 | -0.02 | -0.02 | -0.01 | -0.02 | -0.01 | -0.02 | -0.15 | -0.02 | -0.01 | -0.05 | -0.03 | 1 | | | | | | | | | | |
| 17 | Exporting activity | 0.20 | 0.00 | 0.23 | 0.19 | t=4.01*** | 0.16 | 0.16 | 0.11 | 0.15 | 0.13 | 0.12 | 0.08 | 0.04 | 0.18 | 0.13 | -0.16 | -0.07 | -0.08 | 0.11 | 0.14 | -0.03 | 1 | | | | | | | | | |
| 18 | International quality | 0.27 | 0.01 | 0.30 | 0.25 | t=4.36*** | 0.15 | 0.15 | 0.12 | 0.13 | 0.12 | 0.13 | 0.09 | 0.04 | 0.17 | 0.13 | -0.18 | -0.10 | -0.08 | 0.11 | 0.10 | -0.01 | 0.28 | 1 | | | | | | | | |
| 19 | Technology licenses | 0.17 | 0.00 | 0.17 | 0.17 | t=0.17 | 0.15 | 0.14 | 0.12 | 0.13 | 0.12 | 0.11 | 0.10 | 0.00 | 0.14 | 0.07 | -0.12 | -0.01 | -0.08 | 0.03 | 0.08 | -0.01 | 0.12 | 0.21 | 1 | | | | | | | |
| 20 | Private firm/JV | 0.87 | 0.00 | 0.81 | 0.90 | t=-9.80*** | -0.02 | -0.02 | -0.02 | 0.00 | -0.03 | -0.03 | 0.00 | -0.03 | -0.02 | 0.03 | 0.13 | 0.08 | 0.01 | -0.05 | -0.02 | -0.27 | -0.01 | -0.04 | 0.02 | 1 | | | | | | |
| 21 | Part of larger business | 0.09 | 0.00 | 0.09 | 0.08 | t=1.46 | 0.05 | 0.02 | 0.07 | 0.01 | 0.01 | 0.07 | 0.05 | -0.01 | 0.05 | 0.03 | -0.17 | 0.00 | -0.02 | 0.02 | 0.13 | 0.04 | 0.12 | 0.15 | 0.11 | -0.04 | 1 | | | | | |
| 22 | Public company | 0.86 | 0.00 | 0.90 | 0.84 | t=5.51*** | 0.03 | 0.02 | 0.03 | 0.02 | 0.02 | 0.04 | 0.01 | 0.04 | 0.05 | 0.03 | -0.06 | 0.01 | -0.04 | -0.03 | 0.07 | 0.01 | 0.08 | 0.11 | 0.06 | -0.05 | 0.04 | 1 | | | | |
| 23 | Metropolitan area | 0.35 | 0.01 | 0.34 | 0.35 | t=-0.91 | -0.02 | -0.03 | 0.00 | -0.03 | -0.03 | 0.02 | -0.01 | -0.01 | -0.01 | -0.06 | -0.03 | 0.06 | 0.00 | 0.00 | 0.04 | -0.06 | 0.00 | 0.01 | 0.09 | 0.07 | 0.02 | 0.10 | 1 | | | |
| 24 | Cntry's business dens. | 4.29 | 0.04 | 4.68 | 4.17 | t=5.11*** | 0.12 | 0.10 | 0.11 | 0.08 | 0.09 | 0.10 | 0.09 | 0.07 | 0.05 | 0.09 | 0.05 | -0.07 | 0.10 | 0.06 | 0.04 | -0.09 | 0.02 | 0.02 | -0.05 | 0.09 | -0.09 | 0.16 | -0.10 | 1 | | |
| 25 | Cntry's p. cap. GDP | -0.13 | 0.01 | -0.05 | -0.15 | t=3.24*** | 0.02 | 0.05 | -0.02 | 0.07 | 0.01 | -0.02 | -0.01 | 0.06 | 0.09 | 0.12 | 0.00 | -0.13 | 0.01 | 0.19 | 0.05 | -0.09 | 0.24 | 0.12 | 0.02 | 0.13 | 0.02 | -0.06 | -0.15 | 0.27 | 1 | |
| 26 | Cntry's R&D exp. | -0.29 | 0.01 | -0.15 | -0.33 | t=6.30*** | 0.02 | 0.04 | 0.00 | 0.06 | 0.01 | 0.00 | 0.00 | 0.08 | 0.10 | 0.11 | -0.02 | -0.12 | 0.01 | 0.16 | 0.05 | -0.05 | 0.24 | 0.10 | 0.02 | 0.07 | 0.03 | -0.06 | -0.09 | 0.26 | 0.85 | 1 |

Notes: correlations above |.01| are statistically significant, $p \leq .05$; [a] 16 industry dummies are exclusive. Hence, correlations are not reported. The industry distribution of innovation is statistically significant (overall $\chi^2$= 155.80***) with details as follows: IT (45%), machinery and equipment (38%), electronics (37%), plastics and rubber (36%), other manufacturing (31%), food (31%), basic metals and furniture (30%), wholesale (26%), non-metallic minerals (25%), chemicals (24%), textiles and garment (23%), construction (22%), retail (20%), other services (19%), transport (18%), and hotel and restaurants (17%).

**Table 2: Results of multilevel regression analyses for Hypothesis 1 through 3**

| Multilevel model: | GLS | | | GLS | | | GLS | | |
|---|---|---|---|---|---|---|---|---|---|
| Dependent variable: | Overall Firm Innovativeness | | | Product and Process Innovation | | | Organizational and Marketing Innovation | | |
| Model: | 1a | 1b | 1c | 2a | 2b | 2c | 3a | 3b | 3c |
| *Independent variable* | | | | | | | | | |
| Gender diversity | 0.171*** | 0.144*** | 0.089* | 0.076*** | 0.057** | 0.031 | 0.094*** | 0.084*** | 0.055* |
| in firm ownership | (0.04) | (0.04) | (0.04) | (0.02) | (0.02) | (0.02) | (0.02) | (0.02) | (0.02) |
| *Mediators* | | | | | | | | | |
| R&D investments | | | 1.362*** | | | 0.677*** | | | 0.680*** |
| | | | (0.04) | | | (0.02) | | | (0.02) |
| Breadth of external capital | | | 0.160*** | | | 0.067*** | | | 0.093*** |
| | | | (0.02) | | | (0.01) | | | (0.01) |
| *Control variables* | | | | | | | | | |
| SME | | -0.338*** | -0.273*** | | -0.068 | -0.037 | | -0.265*** | -0.230*** |
| | | (0.07) | (0.06) | | (0.04) | (0.04) | | (0.04) | (0.04) |
| Start-up | | -0.056 | -0.045 | | -0.027 | -0.023 | | -0.029 | -0.023 |
| | | (0.04) | (0.04) | | (0.02) | (0.02) | | (0.02) | (0.02) |
| Top Manager: Female | | -0.020 | 0.002 | | -0.000 | 0.010 | | -0.019 | -0.008 |
| | | (0.04) | (0.03) | | (0.02) | (0.02) | | (0.02) | (0.02) |
| Years of industry experience | | 0.002 | 0.001 | | 0.001 | 0.001 | | 0.001 | 0.001 |
| | | (0.00) | (0.00) | | (0.00) | (0.00) | | (0.00) | (0.00) |
| Foreign ownership | | -0.004 | 0.043 | | -0.035 | -0.014 | | 0.031 | 0.057 |
| | | (0.06) | (0.06) | | (0.04) | (0.03) | | (0.04) | (0.04) |
| State ownership | | -0.077 | -0.118 | | -0.026 | -0.045 | | -0.054 | -0.075 |
| | | (0.13) | (0.12) | | (0.07) | (0.07) | | (0.07) | (0.07) |
| Exporting activity | | 0.322*** | 0.195*** | | 0.185*** | 0.125*** | | 0.136*** | 0.071*** |
| | | (0.04) | (0.04) | | (0.02) | (0.02) | | (0.02) | (0.02) |
| Intern. quality certification | | 0.252*** | 0.146*** | | 0.129*** | 0.078*** | | 0.123*** | 0.069*** |
| | | (0.03) | (0.03) | | (0.02) | (0.02) | | (0.02) | (0.02) |
| Technology license | | 0.430*** | 0.302*** | | 0.230*** | 0.167*** | | 0.198*** | 0.133*** |
| | | (0.04) | (0.03) | | (0.02) | (0.02) | | (0.02) | (0.02) |
| Private firm/JV | | 0.009 | 0.030 | | -0.004 | 0.007 | | 0.011 | 0.021 |
| | | (0.04) | (0.04) | | (0.02) | (0.02) | | (0.03) | (0.02) |
| Part of larger business | | 0.100* | 0.072 | | -0.001 | -0.014 | | 0.107*** | 0.093*** |
| | | (0.05) | (0.05) | | (0.03) | (0.03) | | (0.03) | (0.03) |
| Public company | | 0.019 | -0.019 | | -0.005 | -0.023 | | 0.025 | 0.005 |
| | | (0.04) | (0.04) | | (0.02) | (0.02) | | (0.02) | (0.02) |
| 16 Industry dummies | No | Yes | Yes | No | Yes | Yes | No | Yes | Yes |
| *Overall model statistics* | | | | | | | | | |
| Constant | 0.874*** | 1.129*** | 0.871*** | 0.436** | 0.546** | 0.428** | 0.437** | 0.577*** | 0.438*** |
| | (0.09) | (0.18) | (0.16) | (0.04) | (0.09) | (0.08) | (0.05) | (0.10) | (0.09) |
| ICC[1] | | | 0.126*** | | | 0.086** | | | 0.110*** |
| | | | (0.03) | | | (0.02) | | | (0.03) |
| Snijders/Bosker $R^2$ level (1) | 0.0038 | 0.0719 | 0.2127 | 0.0026 | 0.0729 | 0.1811 | 0.0032 | 0.0558 | 0.1653 |
| Snijders/Bosker $R^2$ level (2) | 0.0124 | 0.0317 | 0.2624 | 0.0130 | 0.0361 | 0.2743 | 0.0103 | 0.0502 | 0.2491 |
| No. of cases level 1 | 7614 | 7614 | 7614 | 7636 | 7637 | 7637 | 7639 | 7639 | 7639 |
| No. of cases level 2 | 29 | 29 | 29 | 29 | 29 | 29 | 29 | 29 | 29 |

*Notes*: standard errors in parentheses. *** $p \leq .001$, ** $p \leq .01$, * $p \leq .05$ (two-tailed tests)
[1]ICC=intra-class correlation

**Table 3. KHB Mediation Analysis Results for Hypotheses 2 and 3**

| Dependent Variable:<br>Model: | Overall Firm Innovativeness<br>(1) | Product and Process Innovation<br>(2) | Organizational and Marketing Innovation<br>(3) |
|---|---|---|---|
| *Summary of effects for specified predictor X on firm innovation* | | | |
| Total direct and indirect effect | 0.16*** | 0.063*** | 0.095*** |
|  | (0.04) | (0.02) | (0.02) |
| Direct effect | 0.10** | 0.034 | 0.063*** |
|  | (0.04) | (0.02) | (0.02) |
| Combined indirect effect | 0.059** | 0.029** | 0.031*** |
|  | (0.02) | (0.01) | (0.01) |
| Total amount mediated | 37.06% | 45.43% | 33.02% |
| *Indirect effects of specified predictor X on firm innovation through proposed mediator* | | | |
| R&D investments | 0.043* | 0.021* | 0.022* |
|  | (0.02) | (0.01) | (0.01) |
|  | 26.84% | 34.01% | 23.09% |
| Breadth of external capital | 0.016*** | 0.007*** | 0.009*** |
|  | (0.00) | (0.00) | (0.00) |
|  | 10.23% | 11.42% | 9.94% |

*Notes*: unstandardized coefficients are displayed in the first row, standard errors in parentheses in the second row, and percentage reduced due to mediation in the third row; all control variables from table 2 are included in the mediation models. *** $p \leq .001$, ** $p \leq .01$, * $p \leq .05$ (two-tailed tests)

**Table 4: HLM Robustness Checks and Mediation for Separate Types of Innovation**

|  | New product | | New processes | | Organizational | | Marketing innovation | |
|---|---|---|---|---|---|---|---|---|
| Independent Variable: Gender diversity in firm | 0.176* (0.08) | 0.102 (0.09) | 0.180* (0.09) | 0.104 (0.09) | 0.280** (0.09) | 0.210* (0.09) | 0.240*** (0.08) | 0.165 (0.09) |
| Mediator 1: R&D investments |  | 1.707*** (0.09) |  | 1.669*** (0.09) |  | 1.712*** (0.09) |  | 1.707*** (0.09) |
| Mediator 2: Breadth of external |  | 0.226*** (0.04) |  | 0.248*** (0.04) |  | 0.302*** (0.04) |  | 0.325*** (0.04) |
| Control variables | all included | all included | all included | all included | all included | all included | all included | all included |
| Total amount of mediation in | n/a | 45.83% | n/a | 35.85% | n/a | 28.21% | n/a | 29.38% |
| McKelvey & Zavoina's $R^2$ | 0.1016 | 0.1681 | 0.0916 | 0.1535 | 0.0894 | 0.1563 | 0.0557 | 0.1269 |
| N of cases level1; level 2 | 7,651; 29 | 7,651; 29 | 7,651; 29 | 7,651; 29 | 7,653; 29 | 7,653; 29 | 7,651; 29 | 7,651; 29 |

Notes: Full results available upon request. *** $p \leq .001$, ** $p \leq .01$, * $p \leq .05$ (two-tailed tests)

**Table 5: Additional HLM Robustness Check and Mediation Results for Hypotheses 1-3**

|  | Model 1: Alternate IV[2] | Model 2: Alternate for M1[3] | Model 3a: Alternate (1) for M2[4a] | Model 3b: Alternate (2) for M2[4b] | Model 4: Without solo-owned firms[5] | Model 5: Without case outliers[6] | Model 6: Higher-innovation industries[7] | Model 7: Lower-innovation industries[8] | Model 8: Heckman two-part model with HLM[9] |
|---|---|---|---|---|---|---|---|---|---|
| IV: Gender diversity in firm ownership[1] | [0.142***] 0.071* | [0.176***] 0.103** | [0.171***] 0.087* | [0.164***] 0.093** | [0.161***] 0.089* | [0.171***] 0.089* | [0.224***] 0.116 | [0.139**] 0.083 | 0.090* n/a |
| M1: R&D | 1.363*** | 0.989*** | 1.361*** | 1.371*** | 1.360*** | 1.362*** | 1.374*** | 1.359*** | n/a |
| M2: External capital | 0.160*** | 0.127*** | 0.235*** | 0.289*** | 0.158*** | 0.160*** | 0.175*** | 0.145*** | 0.111*** |
| Control variables | All included | All included | All included | All included | All included | All included | All included | All included | All included |
| Total amount mediated | 43.80% | 36.39% | 36.35% | 31.96% | 36.95% | 37.06% | 33.39% | 42.85% | n/a |
| Snijders/Bosker $R^2$ level (1) | 0.2124 | 0.2473 | 0.2127 | 0.2078 | 0.2154 | 0.2127 | 0.2483 | 0.1737 | 0.2612 |
| Snijders/Bosker $R^2$ level (2) | 0.2609 | 0.4141 | 0.2663 | 0.2632 | 0.2716 | 0.2624 | 0.3737 | 0.2001 | 0.5226 |
| N of cases level1; level 2 | 7614; 29 | 7544; 29 | 7617; 29 | 7850; 29 | 6873; 29 | 7614; 29 | 3025; 29 | 4589; 29 | 7279; 28 |

*** $p \leq .001$, ** $p \leq .01$, * $p \leq .05$ (two-tailed tests)

[1] DV=overall innovativeness; Regression coefficients for the "gender diversity in firm ownership" without standard errors are displayed in the table (full results are available upon request): values reported with brackets are the regression coefficients based on the analysis that does not include any mediators or controls (analogous to models 1a, 2a, and 3a in Table 3); values without square brackets are the regression coefficients based on the analysis with all mediators and controls included (analogous with model 1c, 2c, and 3c).

[2] Alternate gender diversity index= coded 1 if the firm has a minimum of 59% gender diversity in the ownership structure (else 0).

[3] Alternate measurement for "R&D investments"=HR-pertinent practices enhancing firm innovativeness. Coded 1 if the firm has given its "employees some time to develop or try out a new approach or new idea about products or services, business process, firm management, or marketing" (else 0).

[4a] Alternate measure (1) for external capital=a two-item formative index derived from questions about whether the company had used any of the following external sources for financing its working capital: (1) "borrowed from banks (private and state-owned)", (2) "borrowed from non-bank financial institutions which include microfinance institutions, credit cooperatives, credit unions, or finance companies". Each was coded dichotomously as 1 = yes, 0 = no. Our alternate index for financial capital is thus a count variable ranging from 0 to 2.

[4b] Alternate measure (2) for external capital= Coded 1 if a firm received any "subsidies from the national, regional, or local governments or European Union sources" over the last three years (else 0).

[5] Without solo-owned firms = analysis excludes all firms with sole proprietorship (N=839).

[6] Without case outliers = The "dfbeta" analysis indicated that there were no individual case outliers.

[7] Higher-innovation industries =Industries with above-population-means for innovation intensity of 0.8680 (machinery and equipment, plastics and rubber, basic metals, furniture, non-metallic mineral products, other manufacturing, electronics and IT, wholesale, and food).

[8] Lower-innovation industries = Industries with below-population-means for innovation intensity 0.8680 (retail, hotel and restaurants, other services, construction, transport, textiles and garments, chemicals).

[9] Heckman two-part model with hierarchical linear modeling (HLM). We used the Heckman (1979) model to correct for a potential self-selection bias into R&D activities. The Heckman correction model is a two-step process. In our case, the first step involved estimating a model for explaining an owner-manager's self-selection into R&D activities. We used firm size, firm age, firm industries, location in a metropolitan area, country-fixed effects, firm subsidies, HR-pertinent innovation enhancement practices, foreign ownership, and state ownership, as the predictor variables. In the second step, we estimated a multilevel model for the focal dependent variable, including the inverse Mill's ratio or Rho (i.e., the predicted probability of self-selection into R&D activities) from the selection equation. See the online appendix for more information.